\documentclass[preprintnumbers,amsmath,amssymb]{revtex4}

\usepackage{dcolumn}
\usepackage{bm}
\usepackage{graphicx,subfigure,xcolor}
\usepackage[normalem]{ulem}

\usepackage{comment}
\usepackage{xcolor}
\usepackage{amsmath}
\usepackage{amssymb}
\usepackage{eucal}
\usepackage{mathrsfs}
\usepackage{amsthm}
\usepackage{epstopdf}
\usepackage{textcomp}
\usepackage{braket}
\usepackage{pdfpages}

\begin{document}

\newcommand{\bea}{\begin{eqnarray}}
\newcommand{\eea}{\end{eqnarray}}
\newcommand{\beq}{\begin{equation}}
\newcommand{\eeq}{\end{equation}}
\newcommand{\nn}{\nonumber}
\def\k{{\vec k}}
\def\x{{\vec x}}
\def\/{\over}

\def\tp{\tau'}
\def\xt{x(\tau)}
\def\xtp{x(\tau')}
\def\xAt{x_A(\tau)}
\def\xAtp{x_A(\tau')}
\def\xBt{x_B(\tau)}
\def\xBtp{x_B(\tau')}
\def\w{\omega}
\def\w0{\omega_0}
\def\wk{\omega_k}
\def\bk{\bf k}
\def\bx{\bf x}
\def\ak{a_{\bk}}
\def\akd{a_{\bk}^{\dagger}}
\def\akj{a_{{\bk}j}}
\def\ekj{{\bf{e}}_{{\bk}j}}
\def\akjd{a_{{\bk} j}^{\dagger}}
\def\g{\mid g \rangle}
\def\gd{\langle g \mid}
\def\e{\mid e \rangle}
\def\ed{\langle e \mid}
\def\a{\mid a \rangle}
\def\ad{\langle a \mid}
\def\b{\mid b \rangle}
\def\bd{\langle b \mid}
\def\x{\bf x}
\def\f{\mbox{\small{\em f}}}
\def\s{\mbox{\small{\em s}}}
\def\vac{\mid 0\rangle}
\def\vacd{\langle 0\mid}
\def\arc{arc}
\def\bmu{\boldsymbol\mu}

\title{Resonance dipole-dipole interaction between two accelerated atoms in the presence of a reflecting plane boundary}

\newcommand{\orcidauthorA}{0000-0003-4046-753X}
\newcommand{\orcidauthorB}{0000-0001-8556-5542}
\newcommand{\orcidauthorC}{0000-0003-4884-0028}

\author{Wenting Zhou$^{1,2,3}$\footnote{zhouwenting@nbu.edu.cn}}
\author{Roberto Passante$^{3,4}$\footnote{roberto.passante@unipa.it}}
\author{Lucia Rizzuto$^{3,4}$\footnote{lucia.rizzuto@unipa.it}}

\affiliation{$^{1}$Center for Nonlinear Science and Department of Physics, Ningbo
University, Ningbo, Zhejiang 315211, China}
\affiliation{$^2$China Key Laboratory of Low Dimensional Quantum Structures and Quantum Control of Ministry of Education, Hunan Normal University, Changsha,
Hunan 410081, China}
\affiliation{$^3$Dipartimento di Fisica e Chimica, Universit\`{a} degli Studi di Palermo, Via Archirafi 36, I-90123 Palermo, Italy}
\affiliation{$^4$INFN, Laboratori Nazionali del Sud, I-95123 Catania, Italy}

\begin{abstract}
We study the resonant dipole--dipole interaction energy between two non-inertial identical atoms, one excited and the other in the ground state, prepared in a correlated {\em Bell-type} state, and interacting with the scalar field or the electromagnetic field nearby a perfectly reflecting plate. We suppose the two atoms move with the same uniform acceleration, parallel to the plane boundary, and that their separation is constant during the motion. By separating the contributions of radiation reaction field and vacuum fluctuations to the resonance energy shift of the two-atom system, we show that Unruh thermal fluctuations do not affect the resonance interaction, which is exclusively related to the radiation reaction field. However, non-thermal effects of acceleration in the radiation-reaction contribution, beyond the Unruh acceleration--temperature equivalence, affect the resonance interaction energy.  By considering specific geometric configurations of the two-atom system relative to the plate, we show that the presence of the mirror significantly modifies the resonance interaction energy between the two accelerated atoms.  In particular, we find that new and different features appear with respect to the case of atoms in the free-space, related to the presence of the boundary and to the peculiar structure of the quantum electromagnetic field vacuum in the locally inertial frame. Our~results suggest the possibility to exploit the resonance interaction between accelerated atoms as a probe for detecting the elusive effects of atomic acceleration on radiative~processes.
\end{abstract}
\maketitle

\section{Introduction}

Quantum field theory in accelerated backgrounds has led to deep insights into the fundamental notions of {\em vacuum} and {\em particles}, forcing us to reconsider these basic concepts as observer-dependent notions. A prominent example of this feature is given by the Unruh effect \cite{Unruh1976},  affirming that an observer moving with constant acceleration in the Minkowski vacuum feels a {\em thermal bath} at an Unruh temperature proportional to its proper acceleration, $a$:
\begin{eqnarray}
\label{eq:1}
T_U=\frac{\hbar}{2\pi k_B c} a,
\end{eqnarray}
where $c$ is the speed of light, $\hbar$ the Planck constant, and $k_B$ is the Boltzmann constant.

An analogous effect, in a curved space-time, is the Hawking radiation from a black hole: a~free-falling observer outside a black hole should experience a bath of thermal radiation at the temperature $T_H=\hbar g /(2\pi k_B c)$, $g$ being the local acceleration due to gravity at the event horizon~\cite{Hawking1975}.

As paradoxical as the concept of thermal radiation from vacuum may appear, the Unruh effect is a clear manifestation of the {\em non-unicity} of the notion of quantum vacuum (and of particles), as extensively discussed in the seminal paper by Fulling \cite{Fulling1973} and in following papers on the subject \cite{Davies1975,Unruh1984}.
This conceptually subtle effect, merging classical general relativity and quantum field theory, has been the object of intense investigations in the literature, with different and sometimes conflicting conclusions on its physical interpretation \cite{Crispino2008, Buchholz2013, Rosu2001, Raine1991, Padmanabhan1990, Narozhny2001, Ford2006}.
Additionally,  from Equation (\ref{eq:1}) (cgs units), we have
\begin{eqnarray}
\label{eq:2}
T_U\sim \left( 10^{-23} a \right) \,\,\mbox{K},
\end{eqnarray}
and therefore extremely high accelerations, of the order of $10^{23}~\mbox{cm/s}^{2}$, are necessary to obtain an Unruh thermal bath of a few kelvin, thus making the detection of this effect in the laboratory drastically difficult \cite{Rosu2001, Crispino2008, Schutzhold2006, Retzker2008, Vanzella2001, Martine2011, Pena2014, Cozzella2017}.
Whilst the absence of any experimental observation of the Unruh effect has led to question the reality of the effect  \cite{Ford2006}, it has been argued that the Unruh effect is a fundamental requirement to ensure the consistency of quantum field theory \cite{Matsas2002}. In any case, a direct verification of the effect, and in general of acceleration-dependent effects, could allow us to solve some fundamental controversies about its physical interpretation.

Recently, the effects of an accelerated motion  on the  radiative properties of atoms/molecules in vacuum have been discussed in the literature \cite{Audretsch1994, Audretsch1995, Passante1998, Zhu2006, Rizzuto2011, Rizzuto2016, Lattuca2017}. Changes in the  spontaneous emission rate~\cite{Audretsch1994, Yu2006, Zhu2007, Zhou2012} or in the Lamb shift of single uniformly accelerating atoms \cite{Audretsch1995, Passante1998}, as well as the dispersion Casimir--Polder interaction between a uniformly accelerated atom and a reflecting plate~\cite{Rizzuto2007, Rizzuto2009a, Rizzuto2009b, Zhu2010, She2010} or between two uniformly accelerated atoms \cite{Noto2013, Marino2014}, have been investigated, and their relation with the Unruh effect was discussed.
The effect of non-equilibrium boundaries on radiative properties of atoms has been also considered \cite{Antezza14, Bagarello15}. 

Another, albeit related, problem, recently addressed in the literature, concerns the equivalence between acceleration and temperature. For example, it has been discussed that non-thermal features (related to a uniform acceleration) manifest  in the dispersion (van der Waals/Casimir-Polder) and resonance  interaction between non inertial atoms in the free-space \cite{Marino2014, Rizzuto2016, Zhou2016, Lattuca2017}. These investigations reveal that the effects of a uniform acceleration are not always equivalent to Unruh thermal effects. 

Motivated by these issues, in this paper, we investigate the effect of a non-inertial motion on the resonance interaction between two atoms, that accelerate with the same constant acceleration, parallel to  a reflecting plate. The imposition of boundary conditions on the quantum field on the plate changes vacuum field fluctuations and the density of states of the quantized radiation field, and, thus, it can significantly influence radiative properties of atoms placed nearby \cite{Power1982, Meschede1990, Spagnolo2006, Passante2007, Palacino2017, Zhou2018}.
Our aim is to investigate in detail physical manifestations of atomic acceleration in the radiation-mediated resonance interaction between the two atoms located in the proximity of a reflecting plate.

Resonance and dispersion Casimir--Polder interactions are long-range interactions involving neutral objects such as atoms or molecules \cite{Casimir1948,Salam10}, due to the zero-point fluctuations of the quantum electromagnetic field or to the source field \cite{Compagno1995,Salam10,Rizzuto2007a}. When one or more atoms are in their excited state, a resonance interaction between the atoms can occur, as a result of the exchange of real photons between them. If the two atoms are prepared in a factorized state, the resonance interaction is a fourth-order effect in the coupling and scales as $R^{-2}$ in the far-zone limit, $R\gg\lambda$ ($\lambda$ is the wavelength associated to the main atomic transition, and $R$ is the interatomic distance) \cite{Rizzuto2004}. These interactions, for atoms in a factorized state, have been recently investigated in the literature, also in connection  with some controversial results concerning the presence or not of space oscillating terms \cite{Berman2015, Donaire2015, Barcellona2016, Milonni2015}. Recent results show that the force on the excited state is oscillatory in space, while that on the ground state is monotonic \cite{Donaire2015, Barcellona2016}. A different physical phenomenon occurs if two identical atoms are prepared in a superradiant (or subradiant) Dicke-state. In this case, the resonance interaction energy is obtained at the second-order in the coupling, and it shows space oscillations in the far-zone limit. Such interaction is usually stronger than dispersion interactions and scales as $R^{-1}$, for very large separations ($R\gg\lambda$). Resonance interactions, and the related F\"{o}rster energy transfer \cite{Forster1965}, have been extensively investigated in the literature \cite{Juzelinuas2000}.
The possibility to manipulate (enhance or inhibit) the dispersion and resonance interactions through a structured environment has been also recently investigated \cite{Kurizki1996,  Agarwal1998, Shahmoon2013, Incardone2014, Notararigo2018}.

We consider two atoms moving with the same uniform proper acceleration in a direction parallel to a reflecting boundary and interacting with the quantum scalar and the electromagnetic field in the vacuum state.
Following a procedure originally introduced by Dalibard, Dupont-Roc, and~Cohen-Tannoudji \cite{Dalibard1982, Dalibard1984}, we identify the contribution of self reaction and vacuum fluctuations to the resonance energy shift of the two accelerated atoms \cite{Rizzuto2016, Zhou2016, Zhou2018, Menezes2016}. This~approach has been recently used to investigate radiative process of atoms at rest in the presence of a boundary \cite{Menezes2015, Zhou2018} or in a cosmic string spacetime \cite{ZhouYu2018}, and it has been recently generalized to the fourth order  to evaluate the dispersion Casimir--Polder interaction between two atoms accelerating in the vacuum space \cite{Marino2014}.
We~show that only the radiation reaction field (source field) contributes to the interatomic resonance interaction energy, while vacuum field fluctuations do not.  Consequently,  the resonance interaction does not show Unruh {\em thermal}-like terms (which are related to vacuum field fluctuations). However, non-thermal effects of acceleration appear in the source field contribution, which significantly affect the resonance interaction energy between the two accelerated atoms.
To explore these effects, we consider two distinct geometric configurations of the two-atom-plate system: atoms aligned perpendicular or parallel to the plane boundary. We show that the presence of the mirror significantly modifies the character of the resonance interaction energy between the two accelerated atoms. By an appropriate choice of the orientation of the two dipole moments, we show that new effects of atomic acceleration (not~present for atoms at rest) appear, yielding a non-vanishing resonance interaction energy  even for specific configurations in which the interaction for stationary atoms is zero.  This result also suggests new possibilities of observing the effects of a uniform acceleration through a modification of the resonance interatomic interaction between two identical entangled atoms. Thus, our findings could have relevance for a possible detection of the effect of an accelerated motion  in radiation-mediated interactions between non-inertial~atoms. 

The paper is structured as follows. In Section \ref{Sec 2}, we briefly introduce the method used, and~discuss the resonance interaction energy between two accelerating atoms interacting with a massless relativistic scalar field nearby a reflecting mirror. In Section \ref{Sec3}, we extend our investigation for atoms interacting with the vacuum electromagnetic field. Final remarks and conclusions are given in Section~\ref{Sec4}.

Throughout the paper, we adopt units such that $\hbar=c=k_B=1$.

\section{Resonance Interaction between Two Uniformly Accelerating Atoms: The Scalar Field Case}
\label{Sec 2}
We consider two identical atoms, $A$ and $B$, interacting with a massless relativistic scalar field in the vacuum state and in the presence of a perfectly reflecting plate satisfying Dirichlet boundary conditions.
The two atoms are modeled as point-like systems
with two internal energy levels, $\mp \w0/2$, associated with the eigenstates  $\g$ and $\e$,  respectively.
We suppose that the mirror is located at $z=0$ and that the two atoms move in a direction parallel to the mirror,  with the same uniform proper acceleration, perpendicular to their (constant) separation.
The  atom-field Hamiltonian in the multipolar coupling scheme and within the dipole approximation, in the locally inertial frame of the two atoms (comoving frame), is as follows \cite{Craig1998, Compagno1995, Rizzuto2016, Marino2014}:
\bea
\label{eq:3}
H&=&\omega_0\sigma^A_3(\tau)+\omega_0\sigma^B_3(\tau)+\sum_{\mathbf{k}}\omega_{k}a^{\dag}_{\mathbf{k}}a_{\mathbf{k}}{dt\/d\tau}-\lambda \left( \sigma^A_2(\tau)\phi(x_A(\tau))+\sigma^B_2(\tau)\phi(x_B(\tau)) \right) ,
\eea
where $\sigma_3={1\/2}(|e\rangle\langle e|-|g\rangle\langle g|)$ and $\sigma_2={i\/2}(|g\rangle\langle e|-|e\rangle\langle g|)$ are the pseudospin atomic operators, $a^{\dag}_{\mathbf{k}}$~and $a_{\mathbf{k}}$ are the creation and annihilation operators of the scalar field, $\lambda$ is the coupling constant, and~$x_{\xi}(\tau)$($\xi=A,B$) is the trajectory of atom $\xi$ ($\tau$ is the proper time of the atoms); $\phi(x(\tau))$ is the scalar field operator, with Dirichlet boundary conditions at the surface of the plate. Equation~(\ref{eq:3}) is expressed in the comoving  frame of the two atoms, and we use the Heisenberg~representation.

We assume two identical atoms prepared in one of the two correlated, symmetrical (superradiant), or antisymmetrical (subradiant) states ($|\psi_{+}\rangle$ or $|\psi_{-}\rangle$, respectively):
\begin{eqnarray}
 \label{eq:4}
\vert \psi_{\pm}\rangle={1\/\sqrt{2}}(|g_A,e_B\rangle\pm|e_A,g_B\rangle) .
\end{eqnarray}

To investigate the interatomic resonance dipole--dipole interaction energy, we exploit the procedure originally introduced in Refs. \cite{Dalibard1982, Dalibard1984}, allowing to identify the contributions of the source field and vacuum fluctuations to the interaction energy.
As discussed in \cite{Dalibard1982, Dalibard1984, Marino2014, Rizzuto2016}, this leads to the introduction of an effective Hamiltonian that governs the time evolution of the atomic observables, pertaining to atom $A$ ($B$), given by the sum of two terms (similar expressions are obtained for atom $B$, by exchange of $A$ and $B$):
\bea
\label{eq:5}
(H^{eff}_A)_{vf}=-{i\/2}\lambda^2\int^{\tau}_{\tau_0}d\tau'C^F(x_A(\tau),x_A(\tau'))[\sigma_2^{Af}(\tau),\sigma_2^{Af}(\tau')] ,
\eea
\begin{eqnarray}
(H^{eff}_A)_{sr}&=&-{i\/2}\lambda^2\int^{\tau}_{\tau_0}d\tau'\chi^F(x_A(\tau),x_A(\tau'))\{\sigma_2^{Af}(\tau),\sigma_2^{Af}(\tau')\}
-{i\/2}\lambda^2\int^{\tau}_{\tau_0}d\tau' \left[\chi^F(x_A(\tau),x_B(\tau'))\right.\\
&\ & \times \left. \{\sigma_2^{Af}(\tau),\sigma_2^{Bf}(\tau')\}\right]  ,\label{eq:6}
\end{eqnarray}
where the functions $C^F(x_A(\tau),x_A(\tau '))$ and $\chi^F(x_A(\tau),x_A(\tau '))$ are the field statistical function (symmetric correlation function and the linear susceptibility), respectively:
\bea
\label{eq:7}
C^F(x(\tau),x(\tau'))&=&{1\/2}\langle0|\{\phi(x(\tau)),\phi(x(\tau'))\}|0\rangle ,\\
\label{eq:8}
\chi^F(x(\tau),x(\tau'))&=&{1\/2}\langle0|[\phi(x(\tau)),\phi(x(\tau'))]|0\rangle .
\eea

To obtain the contributions of source field and vacuum fluctuations to the energy shift of the system, we take the average values of the effective Hamiltonians $(H^{eff}_{A(B)})_{vf}$ and $(H^{eff}_{A(B)})_{sr}$ on the correlated state (\ref{eq:4}):
\beq
(\delta E_A)_{vf}=-i\lambda^2\int^{\tau}_{\tau_0}d\tau'C^F(x_A(\tau),x_A(\tau'))\chi^A(\tau,\tau') ,
\label{eq:9}
\eeq
and
\begin{equation}
(\delta E_A)_{sr}=-i\lambda^2\int^{\tau}_{\tau_0}d\tau'\chi^F(x_A(\tau),x_A(\tau'))C^A(\tau,\tau')
-i\lambda^2\int^{\tau}_{\tau_0}d\tau'\chi^F(x_A(\tau),x_B(\tau'))C^{AB}(\tau,\tau'),
\label{eq:10}
\end{equation}
 where $\tau_0\rightarrow -\infty$ and $\tau\rightarrow \infty$ are the initial and final times (similar expressions are obtained for atom~$B$);  $\chi^{A(B)}(\tau,\tau`)$ and $C^{A(B)}(\tau,\tau`)$ are respectively the antisymmetric and symmetric statistical functions of atom $A$ ($B$), while $\chi^{AB}(\tau,\tau`)$ and $C^{AB}(\tau,\tau`)$ refer to the collective two-atom system:
\bea
\label{eq:11}
\chi^{AB}(\tau,\tau')&=&{1\/2}\langle\psi_{\pm}|[\sigma_2^{Af}(\tau),\sigma_2^{Bf}(\tau')]|\psi_{\pm}\rangle ,\\
C^{AB}(\tau,\tau')&=&{1\/2}\langle\psi_{\pm}|\{\sigma_2^{Af}(\tau),\sigma_2^{Bf}(\tau')\}|\psi_{\pm}\rangle .
\label{eq:12}
\eea

From expressions above, it is clear that the resonance interaction is entirely due to the source field contribution \cite{Rizzuto2016}. In fact, Equation (\ref{eq:9}) does not depend on the interatomic distance; it only gives the vacuum fluctuations contribution to the Lamb shift of each atom ($A$ or $B$). Hence, this term does not contribute to the resonance force between the atoms. Similar~considerations apply to the first term on the right-hand side of Equation (\ref{eq:10}).
On the contrary, the second term on the right-hand side of Equation (\ref{eq:10}), which  depends on the distance between the two atoms, is the only contribution relevant at the second order to the interatomic interaction energy. Therefore, the interatomic resonant energy shift is obtained as
\bea
\label{eq:14}
\delta E&=&-i\int^{\tau}_{\tau_0}d\tau'\chi^F(x_A(\tau),x_B(\tau'))C^{AB}(\tau,\tau')+(A\rightleftharpoons B) .
\eea

This conclusion is indeed expected on a physical ground, as the resonance interaction is due to the exchange of a (real and virtual) scalar quantum between the two correlated atoms. It~is thus related to the field emitted by the two atoms (source field). This property has important consequences when we consider the interaction between accelerated atoms. In fact, as discussed in \cite{Rizzuto2016, Lattuca2017}, this interaction energy does not show signatures of the Unruh thermal effect (which is exclusively related to  the vacuum field correlations  in the locally inertial frame). However, we find that the atomic acceleration can determine a qualitative change of the interaction between the two atoms, even if not equivalent to a thermal effect.

We now apply the procedure discussed above to evaluate the resonance interaction energy between two atoms  moving with uniform acceleration, interacting with the vacuum scalar field nearby a reflecting plate. We first evaluate the field's linear susceptibility.  In the presence of a reflecting boundary, it can be expressed as the sum of two terms, a free term ($\chi^F_0$) that coincides with that obtained in free-space, and a boundary-dependent term ($\chi^F_b$), related to the presence of the reflecting plate \cite{Takagi1988}:
\beq
\label{eq:15}
\chi^F(x_A(\tau),x_B(\tau'))=\chi_0^F(x_A(\tau),x_B(\tau'))+\chi_b^F(x_A(\tau),x_B(\tau')) ,
\eeq
with
\bea
\label{eq:16}
\chi_0^F(x_A(\tau),x_B(\tau'))&=&{i\/8\pi|\Delta\mathbf{x}_-|}[\delta(\Delta t+|\Delta\mathbf{x}_-|)
-\delta(\Delta t-|\Delta\mathbf{x}_-|)] ,\\
\label{eq:17}
\chi_b^F(x_A(\tau),x_B(\tau'))&=&{i\/8\pi|\Delta\mathbf{x}_+|}[\delta(\Delta t+|\Delta\mathbf{x}_+|)
-\delta(\Delta t-|\Delta\mathbf{x}_+|)] ,
\eea
where $x_A(\tau)=(t,x,y,z)$, $x_B(\tau')=(t',x',y',z')$, $\Delta t=t-t'$, and $|\Delta\mathbf{x}_{\mp}|=[(x-x')^2+(y-y')^2+(z\mp z')^2]^{1/2}$.

The atomic statistical function $C^{AB}(\tau,\tau`)$ can also be easily obtained \cite{Rizzuto2016}:
\beq
C^{AB}(\tau,\tau')=\pm {1\/8}(e^{i\omega_0(\tau-\tau')}+e^{-i\omega_0(\tau-\tau')})\;,\label{eq:13}
\eeq
where the $\pm$ sign respectively refers to the symmetric or antisymmetric states (Equation (\ref{eq:4})).

Equation (\ref{eq:15}) has a general validity and can be applied to different situations, for example, two atoms at rest in the presence of a mirror or uniformly accelerating near a plane boundary, provided the appropriate atomic trajectories, $x_A(\tau)$ and $x_B(\tau)$, are given.

We now specialize our considerations to two specific cases.
We suppose a mirror located at $z=0$ and assume that the two atoms accelerate in the
half-space $z>0$, with the same uniform proper acceleration, parallel to the reflecting plate. The distance between the atoms is thus constant.
We consider two different geometric configurations of the two-atom system relative to the plate: two atoms aligned along the $z$-axis, perpendicular to the boundary, and two atoms aligned in a direction parallel to the plate. This permits us to simplify our calculation and to discuss some relevant effects of the presence of the plate on the resonant interaction energy between the two accelerating~atoms.

\begin{figure}[h]
\centering
\includegraphics[scale=0.5]{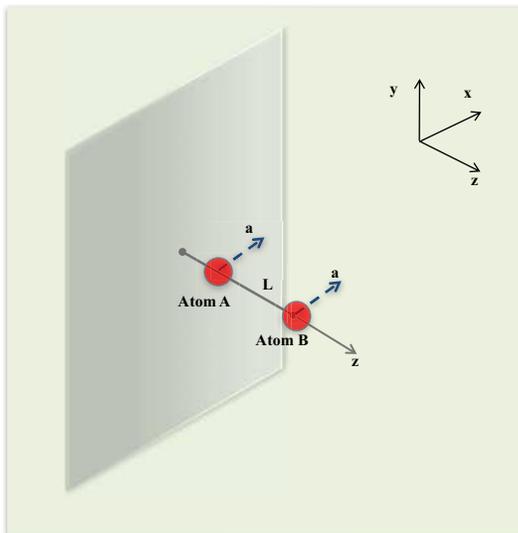}
\caption{Pictorial description of the first geometrical configuration considered for the physical system: two atoms placed on the $z$-axis, perpendicular to the plate, and uniformly accelerating along the~$x$-direction.}
\label{Fig1}
\end{figure}

We first consider both atoms located along the $z$-direction,  perpendicular to the mirror,  and~uniformly accelerating along the $x$-direction, perpendicular to their (constant) separation, as~shown in Figure \ref{Fig1}.

In the locally inertial frame of the two-atom system, the atomic trajectories, as a function of the proper time $\tau$ of both atoms, are 
\begin{eqnarray}
\label{eq:18}
&\ & t_A(\tau)=t_B(\tau)={1\/a}\sinh(a\tau),\,\,\, x_A(\tau)=x_B(\tau)={1\/a}\cosh(a\tau) ,\\
&\ & y_A=y_B=0,\;\,\,\, z_A=z,\,z_B=z+L\, .
\end{eqnarray}

In order to obtain the distance-dependent energy shift of the two-atom system, we first give the linear susceptibility of the scalar field on the  trajectories (Equation (\ref{eq:18})) of the two atoms.
Substituting Equation (\ref{eq:18}) into the expressions of the scalar-field linear susceptibility (Equations  (\ref{eq:16}) and (\ref{eq:17})), we~obtain

\begin{equation}
\label{eq:19}
\chi^F_\perp (x_A(\tau),x_B(\tau'))=-{1\/8\pi^2}\int^{\infty}_0 d\omega(e^{i\omega\Delta\tau}-e^{-i\omega\Delta\tau})
\biggl({\sin({2\omega\/a}\sinh^{-1}({aL\/2}))\/L\sqrt{1+{1\/4}a^2L^2}}
-{\sin({2\omega\/a}\sinh^{-1}({a\mathcal{R}\/2}))\/\mathcal{R}\sqrt{1+{1\/4}a^2\mathcal{R}^2}}\biggr) ,
\end{equation}
where  $\Delta\tau=\tau-\tau`$, $L$ is the interatomic distance, and $\mathcal{R}=z_A+z_B=L+2z$ is the distance between one atom and the image of the second atom relative to the mirror.

The resonance dipole--dipole interaction energy  is then obtained using Equations (\ref{eq:13}) and (\ref{eq:19}) in Equation (\ref{eq:14}). We obtain
\bea
\label{eq:20}
&\ & \delta E_{\perp}(z, L, a) = \mp {\lambda^2\/16\pi}\left[{\cos({2\omega_0\/a}\sinh^{-1}({aL\/2}))\/L\sqrt{1+{1\/4}a^2L^2}}
-{\cos({2\omega_0\/a}\sinh^{-1}({a\mathcal{R}\/2}))\/\mathcal{R}\sqrt{1+{1\/4}a^2\mathcal{R}^2}}\right] ,
\eea
where the $\mp$ sign refers to the symmetric or antisymmetric superposition of the atomic states, respectively.

The expression above describes the resonance dipole--dipole interaction energy in terms of the proper acceleration of the two atoms and the atom-plate distances. In the limit  $a\rightarrow0$, it reduces to that for atoms at rest. It consists of two terms: a term coinciding with the resonance interaction energy for two accelerating atoms in the free-space, discussed in \cite{Rizzuto2016}, and a new term, depending on~$\mathcal{R}$, related to the presence of the mirror. The latter term, describing the effect of the boundary on the energy shift, originates from the interaction of one atom (e.g., atom $A$) with the image of the other atom ($B$). When both atoms are very distant from the reflecting boundary, the boundary-dependent term in Equation (\ref{eq:20}) goes to zero, and we recover the resonance interaction between two atoms accelerating in free-space~\cite{Rizzuto2016}. On the other hand, when the atoms are very close to the mirror, we~can approximate $\mathcal{R}\sim L$, and the resonance interaction is strongly suppressed. Thus, in this limit, the~interaction between the two entangled atoms can be strongly inhibited by means of the nearby plate, analogously to the case of atoms at rest discussed in \cite{Zhou2018}.

Most importantly, Equation (\ref{eq:20}) shows that the effects of the atomic acceleration are not {\em thermal}-like. Nevertheless, the relativistic acceleration significantly affects the interaction energy, giving a different scaling of it with the interatomic distance. In fact, similarly to the results in~\cite{Marino2014, Rizzuto2016} for atoms accelerating in the unbounded space, we can identify a characteristic length scale related to the acceleration,  $z_a=1/a$.  For distances larger than $z_a$, the effects of relativistic acceleration can significantly change the interaction between the two non-inertial atoms; in fact, when $\mathcal{R}>L\gg z_a$, \mbox{we obtain}
\bea
\label{eq:21}
&\ & \delta E_{\perp}(z, L, a) \sim \mp {\lambda^2\/8\pi a}\left[\frac{1}{L^2}{\cos({2\omega_0\/a}\ln({aL\/2}))}
-{\frac{1}{\mathcal{R}^2}\cos({2\omega_0\/a}\ln({a\mathcal{R}\/2}))}\right] ,
\eea
giving a different scaling law of the interaction compared to the case of inertial atoms. 
In the {\em near}-zone limit, $\mathcal{R}, L\ll z_a$, we recover the well-known result for inertial (static) atoms:
\bea
\label{eq:21}
&\ & \delta E_{\perp}(z, L, a) \sim \mp {\lambda^2\/16\pi }\left[\frac{1}{L}{\cos({\omega_0 L})}
-\frac{1}{\mathcal{R}}{\cos({\omega_0 \mathcal{R}})}\right] .
\eea

In the intermediate zone, $\mathcal{R}\gg z_a\gg L$,   when the distance between the two atoms is smaller than the characteristic length $z_a$ but their distance from the mirror is such that $\mathcal{R}\gg z_a$, we obtain
\bea
\label{eq:23}
&\ & \delta E_{\perp}(z, L, a) \sim \mp {\lambda^2\/8\pi }\left[\frac{1}{2L}{\cos({\omega_0 L})}
-{\frac{1}{a\mathcal{R}^2}\cos({2\omega_0\/a}\ln({a\mathcal{R}\/2}))}\right].
\eea

Thus the relativistic acceleration and the presence of the boundary affect the qualitative features of
the resonance interaction, in particular, its power-law distance dependence, decreasing at large distances more rapidly than in the  inertial case.
Additionally, in the presence of a boundary, the~non-inertial character of  acceleration modifies the interatomic interaction energy, even when the separation between the two atoms is much smaller then  $z_a$. In fact, such a result can be expected on a physical ground: the boundary-dependent term, as mentioned, can be interpreted as the interaction of one atom with the image of the other atom with respect to the plate.  
When the atoms are accelerating, the distance traveled by the photon emitted by one atom to reach the other one, after reflection from the mirror, increases with time; if $\mathcal{R} \gg z_a$, this effect becomes relevant and causes an overall decrease of the interaction strength between the two atoms.

\begin{figure}[h]
\centering
\includegraphics[scale=0.5]{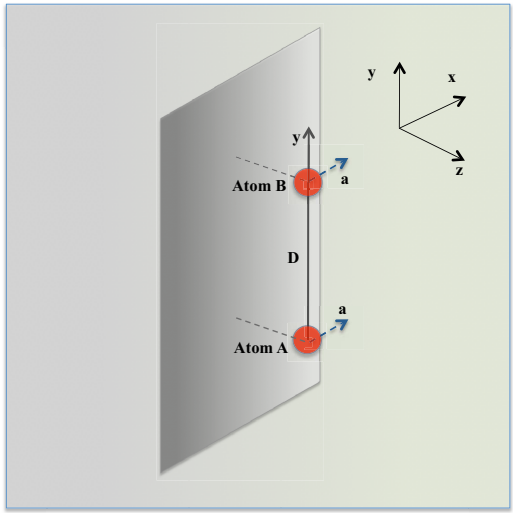}
\caption{Pictorial description of the second geometrical configuration considered for the physical system: two atoms aligned along the $y$-axis, parallel to the plate, and uniformly accelerating along the~$x$-direction.}
\label{Fig2}
\end{figure}

We now investigate whether similar effects manifest also for  a different geometric configuration of the atom-plate system.
Specifically, we consider two atoms aligned in the $y$-direction,  parallel to the mirror, as shown in Figure \ref{Fig2},  and uniformly accelerating in the $x$-direction, perpendicular to their (constant) separation.
In this case, the atomic trajectories are
\begin{eqnarray}
&\ & t_A(\tau)=t_B(\tau)={1\/a}\sinh(a\tau),
\; \; x_A(\tau)=x_B(\tau)={1\/a}\cosh(a\tau),  \\
&\ &y_A=0,\; \; y_B=D,\; \;z_A=z_B=z ,
\label{eq:24}
\end{eqnarray}
with $D>0$.

Following the same procedure as before, we first obtain the scalar-field linear susceptibility:
\begin{equation}
\chi^F_\parallel (x_A(\tau),x_B(\tau'))=-{1\/8\pi^2}\int^{\infty}_0 d\omega(e^{i\omega\Delta\tau}-e^{-i\omega\Delta\tau})
\biggl({\sin({2\omega\/a}\sinh^{-1}({aD\/2}))\/D\sqrt{1+{1\/4}a^2D^2}}
-{\sin({2\omega\/a}\sinh^{-1}({aR\/2}))\/R\sqrt{1+{1\/4}a^2R^2}}\biggr) ,
\label{eq:25}
\end{equation}
where $D$ is the interatomic distance, $\Delta\tau=\tau-\tau`$, and we have defined $R=R(z,D)=\sqrt{D^2+4z^2}$.

The substitution of Equations (\ref{eq:25}) and (\ref{eq:13}) into Equation (\ref{eq:14}) yields, after algebraic calculations, the resonance dipole--dipole interaction for accelerating atoms:
\bea
\delta E_{\parallel}(z, D, a)&=&\mp{\lambda^2\/16\pi}\left[{\cos({2\omega_0\/a}\sinh^{-1}({aD\/2}))\/D\sqrt{1+{1\/4}a^2D^2}}-{\cos({2\omega_0\/a}\sinh^{-1}({aR\/2}))\/R\sqrt{1+{1\/4}a^2R^2}}\right]\;.\label{eq:26}
\eea

As before, we find that the resonance interaction energy consists of two terms. The first term on the right-hand side of Equation (\ref{eq:26}) coincides with that for atoms uniformly accelerating in free-space~\cite{Rizzuto2016}, while the second new term is related to the boundary.  In the static (inertial) limit, we recover the expression of the resonance interaction for atoms at rest near the mirror for the configuration considered~\cite{Zhou2018}:
\bea
\delta E_{\parallel}(z, D)=\mp{\lambda^2\/16\pi}\biggl[{\cos(\omega_0D)\/D}-{\cos(\omega_0\sqrt{D^2+4z^2})\/\sqrt{D^2+4z^2}}\biggr] .
\label{eq:27}
\eea

\begin{figure}[h]
\centering
\includegraphics[scale=0.9]{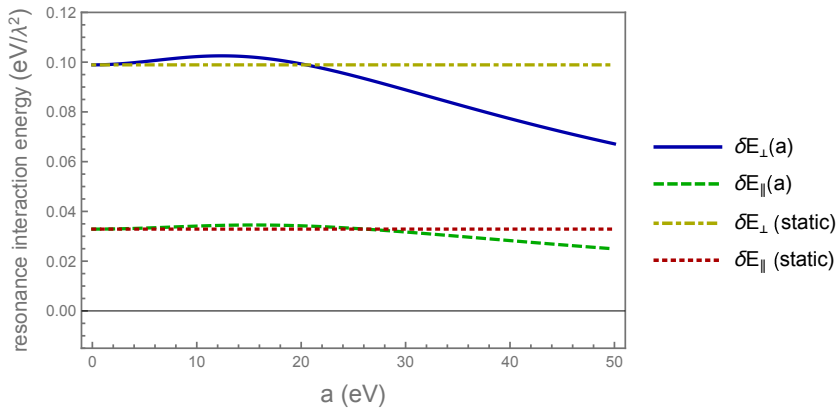}
\caption{Resonance interaction energy  between the two atoms (units: eV$/\lambda^2$, where the coupling constant $\lambda$ in our units is dimensionless), as a function of the atomic acceleration, for two different geometric configurations. Blue continuous line: atoms positioned on the $z$-axis, which is perpendicular to the plate. Green dashed line: atoms along the $y$-axis, which is parallel to the plate. For comparison, the yellow dot-dashed line and the red dotted line respectively refer to the case of inertial atoms aligned in a perpendicular or parallel direction relative to the plate. The plots show that the interaction depends  on the acceleration and on the geometric configuration of the two-atom system relative to the mirror. Parameters, in the units used, are chosen such that $L=D=7.5\times10^{-2}\,$eV$^{-1}$, $z=2.0\times10^{-2}\,$eV$^{-1}$, and $\omega_0=4.17\,$ eV.}
\label{Fig3}
\end{figure}

It is worth noting that the expression of $\delta E_{\parallel}(z, D, a)$  given by Equation (\ref{eq:26}) is formally equal to that obtained for $\delta E_{\perp}(z, L, a)$ in Equation (\ref{eq:20}), provided $\mathcal{R}$ is replaced by $R$.  This is indeed expected, as the distance $R=\sqrt{D^2+4z^2}$ is the distance between one atom and the image of the other. In order to compare the results obtained in the two geometric configurations, in Figure \ref{Fig3} are plotted Equations (\ref{eq:20}) and (\ref{eq:26}) of the resonance interaction energy (in units of eV$/\lambda^2$), as a function of the atomic acceleration. In the plots, the value used for  $\omega_0$ is the ionization energy of $^{87}$Rb, and the distances $L=D$ and $z$ have been chosen in such a way that the plots cover near, intermediate, and far zones, for both perpendicular and parallel alignments of the atoms.
The plots show that the resonance interaction energy depends on the acceleration and the geometric configuration of the two atoms with respect to the plate (perpendicular or parallel alignment) and that it can be enhanced or inhibited, depending on the atomic acceleration.

\section{Resonance Interaction for Two Accelerating Atoms Interacting with the Electromagnetic Field}
\label{Sec3}
In this section, we extend our investigations to  two uniformly accelerated identical atoms interacting with the vacuum electromagnetic field, placed nearby a perfectly reflecting plate. As~before, the atoms move with a uniform proper acceleration $a$ in a direction parallel to the plane, located at $z=0$, and their distance is constant.
Our aim is to discuss whether new  and further effects of acceleration may manifest in their interaction, as a consequence of the vector nature of the electromagnetic field.

We adopt the Hamiltonian in the Coulomb gauge and in the multipolar coupling scheme, within dipole approximation. In the comoving reference frame of both atoms, this is
\begin{equation}
\label{eq:28}
H=\omega_0\sigma^A_3(\tau)+\omega_0\sigma^B_3(\tau)+\sum_{\mathbf{k},\lambda}\omega_{k}a^{\dag}_{\mathbf{k}\lambda}a_{\mathbf{k}\lambda}{dt\/d\tau}
-\bmu_A(\tau)\cdot\mathbf{E}(x_A(\tau))-\bmu_B(\tau)\cdot\mathbf{E}(x_B(\tau)) .
\end{equation}
$\lambda=1,2$ indicates the polarization , $\bmu=e\mathbf{r}$ is the dipole moment operator of the atoms (restricted to the subspace of the two atomic levels considered), and $\mathbf{E}(x(\tau))$ is the electric field operator, with the appropriate boundary conditions on the reflecting plate.

As shown in the previous section, the resonance interaction energy is due only to the radiation-reaction term and can be obtained through the effective Hamiltonian $(H^{eff}_A)_{sr}+(H^{eff}_B)_{sr}$ (terms referring to atoms $A$ and $B$, respectively) on the correlated state  $\vert \psi_{\pm}\rangle$ (see Equations  (\ref{eq:4})--(\ref{eq:6})), taking only terms depending on the interatomic distance:
\begin{equation}
\label{eq:29}
\delta E = -i\int^{\tau}_{\tau_0}d\tau'\chi^F_{ij}(x_A(\tau),x_B(\tau'))C^{AB}_{ij}(\tau,\tau') + (A\rightleftharpoons B),
\end{equation}
where $i,j = x,y,z$. We first evaluate  the electromagnetic field susceptibility $\chi^F_{ij}(x_A(\tau),x_B(\tau`))=\frac 1 2 \langle 0\mid\left [\mathbf{E}_i(x_A(\tau)),\mathbf{E}_j(x_B(\tau`))\right]\mid 0\rangle$ and the atomic symmetric correlation function  $C^{AB}_{ij}(\tau,\tau`)$.

The field susceptibility in the comoving frame can be obtained from the two-point correlation function of the field  \cite{Takagi1988}. The two-point correlation function of the electric field operator in the presence of the reflecting boundary, is the following (for brevity, we omit the time-dependence in the following~expressions):
\beq
\label{eq:30}
g_{ij}(x_A,x_B)=\langle0|E_i(x_A)E_j(x_B)|0\rangle.
\eeq

It can be written as the sum of a free part, $g^{(0)}_{ij}(x_A,x_B)$, and a boundary-dependent term, $g^{(b)}_{ij}(x_A,x_B)$:

\beq
\label{eq:31}
g_{ij}(x_A,x_B)=g^{(0)}_{ij}(x_A,x_B)+g^{(b)}_{ij}(x_A,x_B) ,
\eeq
where
\beq
\label{eq:32}
g^{(0)}_{ij}(x_A,x_B)=-{1\/4\pi^2}(\delta_{ij}\partial_0\partial_{0'}-\partial_i\partial_{j'})
{1\/{(\Delta t-i\epsilon)^2-|\Delta \mathbf{x}_{-}|^2}},
\eeq
\beq
\label{eq:33}
g^{(b)}_{ij}(x_A,x_B)={1\/4\pi^2}[(\delta_{ij}-2n_in_j)\partial_0\partial_{0'}-\partial_i\partial_{j'}]
{1\/{(\Delta t-i\epsilon)^2-|\Delta \mathbf{x}_{+}|^2}} ,
\eeq
and ${n}$ is the unit vector along the line joining the two atoms.

We now specialize our considerations to the two specific configurations considered for the scalar-field case in Section \ref{Sec 2} and illustrated in Figures \ref{Fig1} and \ref{Fig2}, that is, atoms aligned in a direction perpendicular or parallel to the plate, respectively.

\subsection{Atoms Aligned Perpendicularly to the Plate}

We first consider two atoms aligned along the $z$-direction, perpendicular to the boundary, and uniformly accelerating along the $x$-direction, as shown in Figure \ref{Fig1}. Thus they move on the trajectory given by Equation (\ref{eq:18}). Because of the vector structure of the electromagnetic field, the calculation of the field susceptibility turns out to be more complicated than for the scalar field \cite{Takagi1988}. After lengthy algebraic calculations, involving a Lorentz transformation of the fields to the comoving frame, we obtain the following (in the locally inertial frame):
\beq
\label{eq:34}
g_{\perp_{ ij}}(x_A,x_B)=g^{(0)}_{\perp_{ij}}(x_A,x_B)+g^{(b)}_{\perp_{ij}}(x_A,x_B) ,
\eeq
where
\begin{eqnarray}
\label{eq:35}
g^{(0)}_{\perp_{ij}}(x_A,x_B)&=&{a^4\/16\pi^2}{1\/(\sinh^2({a\/2}(\Delta\tau-i\epsilon))-{1\/4}a^2L^2)^3}\times\biggl\{{1\/4}a^2L^2(\delta_{ij}-2n_in_j)\\&&
+\biggl[\delta_{ij}+{1\/2}a^2L^2(\delta_{ij}-k_i k_j-2n_in_j)+a L(k_in_j-k_j n_i)\biggr]\sinh^2\biggl({a\/2}\Delta\tau\biggr)\biggr\}
\end{eqnarray}
 is the two-point correlation function of two atoms uniformly accelerated in vacuum \cite{Rizzuto2016}, and
\begin{eqnarray}
g^{(b)}_{\perp_{ij}}(x_A,x_B)&=&-{a^4\/16\pi^2}{(1-2n_in_j)\/(\sinh^2({a\/2}(\Delta\tau-i\epsilon))-{1\/4}a^2\mathcal{R}^2)^3}\times\biggl\{{1\/4}a^2\mathcal{R}^2(\delta_{ij}-2n_in_j)\\&&
+\biggl[\delta_{ij}+{1\/2}a^2\mathcal{R}^2(\delta_{ij}-k_i k_j-2n_in_j)+a\mathcal{R}(k_in_j+k_j n_i)\biggr]\sinh^2\biggl({a\/2}\Delta\tau\biggr)\biggr\}
\label{eq:36}
\end{eqnarray}
is the contribution due to the presence of the boundary. In the equations above,  ${k}=(1,0,0)$ is a unit vector along the acceleration.
As discussed in \cite{Rizzuto2016}, the function $g_{\perp_{ij}}(x_A,x_B)$ is not isotropic, displaying a non-diagonal component. In fact, in the present case, we have two specific directions in space: the direction perpendicular to the plate and that of the acceleration. Similar anisotropies were already found for a single uniformly accelerated atom near a boundary \cite{Rizzuto2009a} or for two accelerated atoms in the free-space \cite{Rizzuto2016}. They arise from the spatially extended structure of the two-atom-plate system here considered, as well as from the vector character of the electromagnetic field. This peculiarity, as we now show, has deep consequences for the  interaction energy between the two atoms.

In order to evaluate the resonance energy, we first focus our attention on the boundary-dependent term and calculate the linear susceptibility of the electric field. Using Equation (\ref{eq:36}), after lengthly algebraic calculations, involving a Fourier transform of the statistical function of the field, we finally obtain
\begin{eqnarray}
\chi^{F(b)}_{\perp_{ij}}(x_A(\tau),x_B(\tau'))&=&{1\/8\pi^2}\int^{\infty}_0d\omega(e^{i\omega\Delta\tau}-e^{-i\omega\Delta\tau})
\biggl(f^{\perp (b)}_{ij}(a,\mathcal{R},\omega)\cos\biggl({2\omega\/a}\sinh^{-1}\biggl({a\mathcal{R}\/2}\biggr)\biggr)\\&&
+h^{\perp (b)}_{ij}(a,\mathcal{R},\omega)\sin\biggl({2\omega\/a}\sinh^{-1}\biggl({a\mathcal{R}\/2}\biggr)\biggr)\biggr) ,
\label{eq:37}
\end{eqnarray}
where we have introduced the functions $f^{\perp (b)}_{ij}(a,\mathcal{R},\omega)$ and $h^{\perp (b)}_{ij}(a,\mathcal{R},\omega)$ given in Appendix \ref{aa} see Equations  (\ref{f3}) and (\ref{h3})).

Substituting Equation (\ref{eq:37}) and the atomic symmetric statistical function:
\beq
C^{AB}_{ij}(\tau,\tau')={1\/2}(\mu^A_{ge})_i(\mu^B_{ge})_j(e^{i\omega_0\Delta\tau}+e^{-i\omega_0\Delta\tau}) ,
\label{eq:38}
\eeq
into Equation (\ref{eq:29}), we finally obtain the boundary-dependent contribution to the resonant energy shift of the two accelerating atoms:
\begin{equation}
\delta E_{\perp}^{(b)}=\mp{1\/4\pi}[\delta_{ij}(\mu^A_{ge})_i(\mu^B_{eg})_jP^{\perp (b)}_{ij}(a,\mathcal{R},\omega_0)
\pm((\mu^A_{ge})_x(\mu^B_{eg})_z+(\mu^A_{ge})_z(\mu^B_{eg})_x)P^{\perp (b)}_{xz}(a,\mathcal{R},\omega_0)] ,
\label{eq:39}
\end{equation}
where we have introduced the function $P^{\perp (b)}_{ij}(a,\mathcal{R},\omega_0)$:
\begin{equation}
P^{\perp (b)}_{ij}(a,\mathcal{R},\omega_0)=f^{\perp (b)}_{ij}(a,\mathcal{R},\omega_0)\sin\biggl({2\omega_0\/a}\sinh^{-1}\biggl({a\mathcal{R}\/2}\biggr)\biggr)
-h^{\perp (b)}_{ij}(a,\mathcal{R},\omega_0)\cos\biggl({2\omega_0\/a}\sinh^{-1}\biggl({a\mathcal{R}\/2}\biggr)\biggr) ,
\label{eq:40}
\end{equation}
modulating the interaction as a function of $\mathcal{R}$ and of the atomic acceleration.

With a similar procedure, evaluation of the boundary-independent contribution, $\delta E_{\perp}^{(0)}$,  to the resonance interaction energy yields the following \cite{Rizzuto2016}:
\begin{equation}
\delta E_{\perp}^{(0)}=\pm{1\/4\pi}[\delta_{ij}(\mu^A_{ge})_i(\mu^B_{eg})_j\mathrm{P}^{\perp (0)}_{ij}(a,L,\omega_0)
\pm((\mu^A_{ge})_x(\mu^B_{eg})_z-(\mu^A_{ge})_z(\mu^B_{eg})_x)\mathrm{P}^{\perp (0)}_{xz}(a,L,\omega_0)] ,
\label{eq:41}
\end{equation}
where
\begin{equation}
\mathrm{P}^{\perp (0)}_{ij}(a,L,\omega_0)=\mathrm{f}^{\perp (0)}_{ij}(a,L,\omega_0)\sin\biggl({2\omega_0\/a}\sinh^{-1}\biggl({aL\/2}\biggr)\biggr)
-\mathrm{h}^{\perp (0)}_{ij}(a,L,\omega_0)\cos\biggl({2\omega_0\/a}\sinh^{-1}\biggl({aL\/2}\biggr)\biggr) ,
\label{eq:42}
\end{equation}
and the functions $\mathrm{f}^{\perp (0)}_{ij}(a,L,\omega)$ and $\mathrm{h}^{\perp (0)}_{ij}(a,L,\omega)$ are given by Equations  (\ref{tilderf3}) and (\ref{tilderh3}) of Appendix \ref{aa}.

The complete resonance interaction energy of the accelerated two-atom system is then obtained by summing Equations (\ref{eq:39}) and  (\ref{eq:41}):
\bea
&&\delta E_{\perp}=\delta E_{\perp}^{(0)}+\delta E_{\perp}^{(b)} .
\label{eq:43}
\eea

The result (Equation (\ref{eq:43})) is valid for any value of the parameters $a$, $L$, and $\mathcal{R}$. It is easy to show that in the {\em near-zone} limit,  $L\ll a^{-1}$ and $\mathcal{R}\ll a^{-1}$, the linear susceptibility is well described by its stationary counterpart, and we recover the expression of the resonance interaction for two atoms at rest~\cite{Rizzuto2016, Zhou2018}. 
However, at higher orders in $a\mathcal{R}$ (and/or $aL$), corrections related to the accelerated motion of the two atoms become relevant, yielding a different scaling of the interaction energy with the distance, in analogy to the scalar-field case  discussed in the previous section.
Interestingly, a comparison with the scalar-field case shows the emergence of  new features in the resonance interaction, due to the boundary, and related to the anisotropic structure of the electromagnetic field susceptibility.
Indeed, from Equation (\ref{eq:39}), it follows that the effect of the acceleration on the interaction can be controlled by an appropriate choice of the dipoles' orientations and of the distance of the two atoms from the plate.
For example, when the dipole moments are orthogonal to each other, with one along $x$ and the other along $z$, the diagonal term in Equation (\ref{eq:39}) vanishes, and only the second (non-diagonal) term survives. The non-diagonal term is present only for $a\not= 0$, and its contribution is a peculiar characteristic of the non inertial atomic  motion, giving a non-vanishing interaction energy, in a configuration where that for static atoms is zero. This term is thus a sharp signature of an accelerated motion. To numerically estimate this energy shift, we can assume  $a=10^{18}$m/s$^2$ ($2.2\times10^{-6}$ eV, in our units), $z=10^{-8}$ m ($\sim 5\times 10^{-2}$ eV$^{-1}$), $L=1.5\times 10^{-8}$ m ($\sim 7.5\times 10^{-2}$ eV$^{-1}$), and $\hbar\omega_0=4.17$ eV, obtaining  $\delta E\simeq 4.4\times 10^{-10}$ eV.  This energy shift is about 4 orders of magnitude smaller than the Lamb shift for the $n=2$ level of the hydrogen atom. Although quite small, we expect that such an energy shift should be measurable using high-resolution spectroscopy, provided the assumed constant acceleration could be reached.   

The results above suggest investigation of whether analogous effects of acceleration manifest also for other geometric configurations of the two atoms system, for example, when both atoms are aligned parallel to the reflecting plane boundary. This configuration is considered in the next subsection.

\subsection{Atoms Aligned Parallel to the Plate}

 We now consider the configuration of two atoms aligned along the $y$-direction, parallel to the boundary, which move with uniform proper acceleration along the $x$-direction, such that their trajectories are those given by Equation (\ref{eq:24}). As before, the distance between the two atoms  remains constant during their motion. This configuration is illustrated in Figure \ref{Fig2}.

The two-point correlation function of the field in the locally inertial frame of both atoms is
\beq
\label{eq:44}
g_{\parallel_{ij}}(x_A,x_B)=g^{(0)}_{\parallel_{ij}}(x_A,x_B)+g^{(b)}_{\parallel_{ij}}(x_A,x_B)\; ,
\eeq
where $g^{(0)}_{\parallel_{ij}}(x_A,x_B)$ is the two-point correlation function in free-space
\cite{Rizzuto2016} and $g^{(b)}_{\parallel_{ij}}(x_A,x_B)$ is the boundary-dependent contribution, which consists of a diagonal term:
\begin{eqnarray}
\label{eq:45}
g^{(b)}_{\parallel_{ij}}(x_A,x_B)&=&-{a^4\/16\pi^2}{(\delta_{ij}-2n_in_j)\/(\sinh^2({a\/2}(\Delta\tau-i\epsilon))-{1\/4}a^2R^2)^3}
\biggl\{{1\/4}a^2\tilde{R}^2(n_in_j-p_ip_j)\\&&+{1\/4}a^2R^2k_i k_j+\biggl[1+{1\/2}a^2\tilde{R}^2(1-k_i k_j-2p_ip_j)\biggr]
\sinh^2\biggl({a\/2}\Delta\tau\biggr)\biggr\} \, \, \, \, (i=j)\; 
\end{eqnarray}
that is non-vanishing only for $i=j$, and a non-diagonal term:
\begin{eqnarray}
&& g^{(b)}_{\parallel_{ij}}(x_A,x_B)=-{a^4\/16\pi^2}{1\/(\sinh^2({a\/2}(\Delta\tau-i\epsilon))-{1\/4}a^2R^2)^3}
\biggl\{-a^2zD(p_in_j-p_j n_i)\\
&&+[a D(k_ip_j-k_j p_i)+2a z(k_in_j+k_j n_i)-2a^2zD(p_in_j-p_j n_i)]
\sinh^2\biggl({a\/2}\Delta\tau\biggr)\biggr\} \, \, \, \, (i\neq j)
\label{eq:46}
\end{eqnarray}
that is different from zero only for $i\neq j$.  We have here introduced the unit vector $\mathbf{p}=(0,1,0)$ and the distances $R=\sqrt{D^2+4z^2}$ and $\tilde{R}=\sqrt{D^2-4z^2}$). The boundary-dependent contribution to the linear susceptibility of the field is then obtained as
\begin{eqnarray}
\label{eq:47}
\chi^{F(b)}_{\parallel_{ij}}(x_A(\tau),x_B(\tau'))&=&{1\/8\pi^2}\int^{\infty}_0d\omega(e^{i\omega\Delta\tau}-e^{-i\omega\Delta\tau})
\biggl(f^{\parallel(b)}_{ij}(a,D,z,\omega)\cos\biggl({2\omega\/a}\sinh^{-1}\biggl({aR\/2}\biggr)\biggr)\\&&
+h^{\parallel(b)}_{ij}(a,D,z,\omega)\sin\biggl({2\omega\/a}\sinh^{-1}\biggl({aR\/2}\biggr)\biggr)\biggr) ,
\end{eqnarray}
where the functions $f^{\parallel (b)}_{ij}(a,D,z,\omega)$ and $h^{\parallel (b)}_{ij}(a,D,z,\omega)$, given in Equations  (\ref{f4}) and (\ref{h4}) of Appendix~\ref{aa}, modulate the resonance interaction energy with the distance $D$ and the atomic acceleration~$a$.

Substituting Equations  (\ref{eq:47}) and (\ref{eq:38}) into Equation (\ref{eq:29}), we find the boundary-dependent contribution to the resonant energy shift:
\begin{eqnarray}
\delta E_{\parallel}^{(b)}=-{1\/4\pi}\left[ \delta_{ij}(\mu^A_{ge})_i(\mu^B_{eg})_jP^{\parallel (b)}_{ij}(a,D,z,\omega_0)
+\left( (\mu^A_{ge})_x(\mu^B_{eg})_y-(\mu^A_{ge})_y(\mu^B_{eg})_x \right)P^{\parallel (b)}_{xy}(a,D,z,\omega_0) \right. \\
+\left. \! \left( (\mu^A_{ge})_x(\mu^B_{eg})_z+(\mu^A_{ge})_z(\mu^B_{eg})_x\right) P^{\parallel (b)}_{xz}(a,D,z,\omega_0)
+\left( (\mu^A_{ge})_y(\mu^B_{eg})_z-(\mu^A_{ge})_z(\mu^B_{eg})_y\right) P^{\parallel (b)}_{yz}(a,D,z,\omega_0) \right] ,
\label{eq:48}
\end{eqnarray}
where
\begin{eqnarray}
\label{eq:49}
P^{\parallel (b)}_{ij}(a,D,z,\omega_0)&=&f^{\parallel (b)}_{ij}(a,D,z,\omega_0)\sin\biggl({2\omega_0\/a}\sinh^{-1}\biggl({aR\/2}\biggr)\biggr)\\
&-&h^{\parallel (b)}_{ij}(a,D,z,\omega_0)\cos\biggl({2\omega_0\/a}\sinh^{-1}\biggl({aR\/2}\biggr)\biggr).
\end{eqnarray}

The resonance interaction energy between the accelerating atoms is finally obtained by adding Equation (\ref{eq:48}) to the free-space interaction energy  $\delta E_{\parallel}^{(0)}$, given by the following \cite{Rizzuto2016}:
\begin{equation}
\delta E_{\parallel}^{(0)}={1\/4\pi}\left[ \delta_{ij}(\mu^A_{ge})_i(\mu^B_{eg})_j\mathrm{P}^{\parallel (0)}_{ij}(a,D,\omega_0)
+\left( (\mu^A_{ge})_x(\mu^B_{eg})_y-(\mu^A_{ge})_y(\mu^B_{eg})_x\right) \mathrm{P}^{\parallel (0)}_{xy}(a,D,\omega_0) \right]\; ,
\label{eq:50}
\end{equation}
with
\begin{eqnarray}
\label{eq:51}
&&\mathrm{P}^{\parallel (0)}_{ij}(a,D,\omega_0)=\mathrm{f}^{\parallel (0)}_{ij}(a,D,\omega_0)\sin\biggl({2\omega_0\/a}\sinh^{-1}\biggl({aR\/2}\biggr)\biggr)\\&&\quad\quad
-\mathrm{h}^{\parallel (0)}_{ij}(a,D,\omega_0)\cos\biggl({2\omega_0\/a}\sinh^{-1}\biggl({aR\/2}\biggr)\biggr)\;
\end{eqnarray}
(the functions $\mathrm{f}^{\parallel (0)}_{ij}(a,D,\omega)$ and $\mathrm{h}^{\parallel (0)}_{ij}(a,D,\omega)$ can be obtained from Equations  (\ref{tilderf3}) and (\ref{tilderh3}) in Appendix \ref{aa}  by exchanging subscripts $z$ and $y$).

A comparison with the case of accelerated atoms aligned along the $z$-axis, considered in the previous subsection, shows the emergence of a new effect, related to the specific geometric configuration of the two-atom system with respect to the plane boundary. In fact, from the equations above, it follows that when the dipole moments are orthogonal to each other, one of them along $y$ and the other in the plane $xz$, a new  non-vanishing contribution to the interaction energy (not present for atoms located perpendicular to the boundary) arises. This contribution exists only when $a\not= 0$, and thus it is  a peculiarity of an accelerated motion.
This gives new additional possibilities to exploit the resonance interaction between accelerated atoms for detecting (non-thermal) effects of acceleration and, in general, physical effects of the accelerated motion on radiation-mediated interactions between atoms. 

\section{Summary}
\label{Sec4}
We have discussed the resonance energy shift of two identical atoms, one excited and the other in the ground state,  prepared in a correlated (superradiant or subradiant) state, and moving with uniform acceleration near a perfectly reflecting plate.
The atoms interact with the massless scalar field or the electromagnetic field in the vacuum state. Following the approach in Refs. \cite{Dalibard1982, Dalibard1984}, we have identified the contributions of source field and vacuum fluctuations to the resonance interaction. We have shown that Unruh thermal fluctuations do not influence the resonance interatomic interaction, which is  obtained from the source-field term only. We show that, in cases of both the scalar and  electromagnetic field, the presence of the plane boundary significantly affects the resonance interaction between the accelerated atoms. Non-thermal effects of acceleration appear, yielding a change in the distance dependence of the interaction. Finally, in the case of the electromagnetic field, we show, for different configurations of the two-atom-plate system, the emergence of new and different effects in the resonance interaction energy, for example, a non-vanishing interaction energy in configurations/dipole orientations for which the interaction is zero for inertial atoms. These effects, not present for atoms at rest,  therefore provide a sharp signature of the non-inertial motion of the atoms. 
These findings could be exploited for the detection of the non-thermal effects of atomic acceleration in radiation-mediated interactions between non-inertial atoms.

\vspace{6pt}
\begin{acknowledgments}
W.Z. acknowledges financial support from the National Natural Science Foundation of China (NSFC) under Grant Nos. 11405091, 11690034, 11375092, and 11435006; the Key Laboratory of Low-Dimensional Quantum Structures and Quantum Control of Ministry of Education under Grant No. QSQC1801; the China Scholarship Council (CSC); the Research program of Ningbo University under Grant No. XYL18027; and the K. C. Wong Magna Fund of Ningbo University.
R.P. and L.R. gratefully acknowledge financial support from the Julian Schwinger Foundation.
\end{acknowledgments}


\appendix
\section{}\label{aa}
In this Appendix, we give the expressions of the functions $f^{\perp (\parallel)}_{ij}$ and $h^{\perp (\parallel)}_{ij}$ used in Section \ref{Sec3}.

The explicit expressions of the functions $f^{\perp (b)}_{ij}(a,\mathcal{R},\omega)$ and $h^{\perp (b)}_{ij}(a,\mathcal{R},\omega)$ are
\bea
\left\{
  \begin{array}{ll}\vspace{5pt}

    f^{\perp (b)}_{xx}={\omega(1+a^2\mathcal{R}^2)\/\mathcal{N}^4\mathcal{R}^2} ,\\\vspace{5pt}
    f^{\perp (b)}_{yy}={\omega(1+{1\/2}a^2\mathcal{R}^2)\/{\mathcal{N}}^2\mathcal{R}^2} ,\\\vspace{5pt}
    f^{\perp (b)}_{zz}={\omega(2+{1\/4}a^2\mathcal{R}^2+{1\/8}a^4\mathcal{R}^4)\/\mathcal{N}^4\mathcal{R}^2} ,\\\vspace{5pt}
    f^{\perp (b)}_{xz}=f^{\perp (b)}_{zx}=-{a\omega(1-{1\/2}a^2\mathcal{R}^2)\/2\mathcal{N}^4\mathcal{R}} ,
  \end{array}
\right.
\label{f3}
\eea
\bea
\left\{
  \begin{array}{ll}\vspace{5pt}
    h^{\perp (b)}_{xx}=-{1+{1\/2}a^2\mathcal{R}^2+{1\/4}a^4\mathcal{R}^4\/\mathcal{N}^{5}\mathcal{R}^3}+{\omega^2\/{\mathcal{N}}^{3}\mathcal{R}} ,\\\vspace{5pt}
    h^{\perp (b)}_{yy}=-{1\/{\mathcal{N}}^{3}\mathcal{R}^3}+{\omega^2\/\mathcal{N}\mathcal{R}} ,\\\vspace{5pt}
    h^{\perp (b)}_{zz}=-{2(1+{5\/8}a^2{\mathcal{R}}^2)\/\mathcal{N}^{5}\mathcal{R}^3}+{a^2\mathcal{R}\omega^2\/4\mathcal{N}^3} ,\\\vspace{5pt}
    h^{\perp (b)}_{xz}=h^{\perp (b)}_{zx}={a(1+a^2\mathcal{R}^2)\/2\mathcal{N}^5\mathcal{R}^2}+{a\omega^2\/2{\mathcal{N}}^3} ,
  \end{array}
\right.
\label{h3}
\eea
with $\mathcal{N}=\mathcal{N}(a,\mathcal{R})=\sqrt{1+{1\/4}a^2\mathcal{R}^2}$.

Explicit expressions of $\mathrm{f}^{\perp (0)}_{ij}(a,L,\omega)$ and $\mathrm{h}^{\perp (0)}_{ij}(a,L,\omega)$ are
\bea
\left\{
  \begin{array}{ll}\vspace{5pt}
    \mathrm{f}^{\perp (0)}_{xx}={\omega(1+a^2L^2)\/N^4L^2} ,\\\vspace{5pt}
    \mathrm{f}^{\perp (0)}_{yy}={\omega(1+{1\/2}a^2L^2)\/N^2L^2} ,\\\vspace{5pt}
    \mathrm{f}^{\perp (0)}_{zz}=-{\omega(2+{1\/4}a^2L^2+{1\/8}a^4L^4)\/N^4L^2} ,\\\vspace{5pt}
    \mathrm{f}^{\perp (0)}_{xz}=-\mathrm{f}^{\perp (0)}_{zx}={a\omega(1-{1\/2}a^2L^2)\/2N^4L} ,
  \end{array}
\right.
\label{tilderf3}
\eea
\bea
\left\{
  \begin{array}{ll}\vspace{5pt}
    \mathrm{h}^{\perp (0)}_{xx}=-{1+{1\/2}a^2L^2+{1\/4}a^4L^4\/N^{5}L^3}+{\omega^2\/N^{3}L} ,\\\vspace{5pt}
    \mathrm{h}^{\perp (0)}_{yy}=-{1\/N^3L^3}+{\omega^2\/N^{{1/2}}L} ,\\\vspace{5pt}
    \mathrm{h}^{\perp (0)}_{zz}={2(1+{5\/8}a^2L^2)\/N^5L^3}-{a^2L\omega^2\/4N^3} ,\\\vspace{5pt}
    \mathrm{h}^{\perp (0)}_{xz}=-\mathrm{h}^{\perp (0)}_{zx}=-{a(1+a^2L^2)\/2N^{5}L^2}-{a\omega^2\/2N^{3}} ,
  \end{array}
\right.
\label{tilderh3}
\eea
with $N=N(a,L)=\sqrt{1+{1\/4}a^2L^2}$.

Explicit expressions of $f^{\parallel (b)}_{ij}(a,D,z,\omega)$ and $h^{\parallel (b)}_{ij}(a,D,z,\omega)$ are
\bea
\left\{
  \begin{array}{ll}\vspace{5pt}
    f^{\parallel (b)}_{xx}={\omega(1+a^2R^2)\/\tilde{N}^4R^2} ,\\\vspace{5pt}
    f^{\parallel (b)}_{yy}={\omega[4z^2-2D^2-{1\/4}a^2R^2(D^2-12z^2)-{1\/8}a^4R^4(D^2-4z^2)]\/\tilde{N}^4R^4} ,\\\vspace{5pt}
    f^{\parallel (b)}_{zz}={\omega[z^2(16+2a^2R^2+a^4R^4)-D^2(2+{3\/2}a^2R^2+{1\/4}a^4R^4)]\/2\tilde{N}^4R^4},\\\vspace{5pt}
    f^{\parallel (b)}_{xy}= - f^{\parallel (b)}_{yx}=-{\omega aD(1-{1\/2}a^2R^2)\/2\tilde{N}^4R^2} ,\\\vspace{5pt}
    f^{\parallel (b)}_{xz}= f^{\parallel (b)}_{zx}=-{\omega az(1-{1\/2}a^2R^2)\/\tilde{N}^4R^2} ,\\\vspace{5pt}
    f^{\parallel (b)}_{yz}= - f^{\parallel (b)}_{zy}=-{2\omega zD(3+a^2R^2+{1\/4}a^4R^4)\/\tilde{N}^4R^4} ,
  \end{array}
\right.
\label{f4}
\eea
\bea
\left\{
  \begin{array}{ll}\vspace{5pt}
    h^{\parallel (b)}_{xx}=-{1+{1\/2}a^2R^2+{1\/4}a^4R^4\/\tilde{N}^{5}R^3}+{\omega^2\/\tilde{N}^{3}R} ,\\\vspace{5pt}
    h^{\parallel (b)}_{yy}={2D^2-4z^2+{1\/4}a^2R^2(5 D^2-4z^2)\/\tilde{N}^{5}R^5}+{\omega^2[4z^2-{1\/4}a^2R^2(D^2-4z^2)]\/\tilde{N}^3R^3} ,\\\vspace{5pt}
    h^{\parallel (b)}_{zz}={D^2(1+{1\/4}a^2R^2)-8z^2(1+{5\/8}a^2R^2)\/\tilde{N}^{5}R^5}+{\omega^2[a^2z^2R^2- D^2(1+{1\/4}a^2R^2)]\/\tilde{N}^{3}R^3} ,\\\vspace{5pt}
    h^{\parallel (b)}_{xy}=-h^{\parallel (b)}_{yx}={a D(1+a^2R^2)\/2\tilde{N}^{5}R^3}+{\omega^2aD\/2\tilde{N}^{3}R} ,\\\vspace{5pt}
    h^{\parallel (b)}_{xz}=h^{\parallel (b)}_{zx}={a z(1+a^2R^2)\/\tilde{N}^{5}R^3}+{\omega^2az\/\tilde{N}^{3}R} ,\\\vspace{5pt}
    h^{\parallel (b)}_{yz}= - h^{\parallel (b)}_{zy}={6zD(1+{1\/2}a^2R^2)\/\tilde{N}^{5}R^5}-{2\omega^2 z D(1+{1\/2}a^2R^2)\/\tilde{N}^{3}R^3} ,
  \end{array}
\right.
\label{h4}
\eea
with $\tilde{N}=\tilde{N}(a,R)=\sqrt{1+{1\/4}a^2R^2}$.

\baselineskip=16pt



\begin{thebibliography}{99}
\bibitem{Unruh1976} Unruh, W.G. Notes on black-hole evaporation. {\em Phys. Rev. D} {\bf 1976}, {\em 14}, 870, doi:10.1103/PhysRevD.14.870.
\bibitem{Hawking1975} Hawking, S.W. Particle creation by black holes. {\em Commun. Math. Phys.} {\bf 1975}, {\em 43}, 199, doi:10.1007/BF01608497.
\bibitem{Fulling1973} Fulling, S.A. Non-uniqueness of Canonical Field Quantization in Riemanian Space-Time. {\em Phys. Rev. D} {\bf 1973}, {\em 7}, 2850, doi:10.1103/PhysRevD.7.2850.
\bibitem{Davies1975} Davies, P.C.W. Scalar particle production in Schwarzschild and Rindler metrics. {\em J. Phys. A} {\bf 1975}, {\em 8}, 609, doi:10.1088/0305-4470/8/4/022.
\bibitem{Unruh1984} Unruh, W.G.; Wald, R.M. What happens when an accelerating observer detects a Rindler particle. \mbox{{\em Phys. Rev. D}} {\bf 1984}, {\em 29}, 1047, doi:10.1103/PhysRevD.29.1047.
\bibitem{Crispino2008} Crispino, L.C.B.; Higuchi, A.; Matsas, G.E.A. The Unruh effect and its applications. {\em Rev. Mod. Phys.} {\bf 2008}, {\em 80}, 787, doi:10.1103/RevModPhys.80.787.
\bibitem{Buchholz2013} Buchholz, D.; Solveen, C. Unruh effect and the concept of temperature. {\em Class. Quantum Gravity} {\bf 2013}, {\em 30}, 085011, doi:10.1088/0264-9381/30/8/085011.
\bibitem{Rosu2001} Rosu, H.C. Hawking like effects and Unruh like effects: Towards experiments? {\em Grav. Cosmol.} {\bf 2001}, {\em 7}, 1--17.
\bibitem{Raine1991} Raine, D.J.; Sciama, D.W.; Grove, P.G. Does a uniformly accelerated quantum oscillator radiate? {\em Proc. R. Soc. Lond. A} {\bf 1991}, {\em 435}, 205, doi:10.1098/rspa.1991.0139.
\bibitem{Padmanabhan1990} Padmanabhan, T. Physical interpretation of quantum field theory in noninertial coordinate systems. \mbox{{\em Phys. Rev. Lett.}} {\bf 1990}, {\em 64}, 2471, doi:10.1103/PhysRevLett.64.2471.
\bibitem{Narozhny2001} Narozhny, N.B.; Fedotov, A.M.; Karnakov, B.M.; Mur, V.D.; Belinskii, V.A. Boundary conditions in the Unruh problem. {\em Phys. Rev. D} {\bf 2001}, {\em 65}, 025004, doi:10.1103/PhysRevD.65.025004.
\bibitem{Ford2006} Ford, G.W.; O'Connell, R.F. Is there Unruh radiation? {\em Phys. Lett. A} {\bf 2006}, {\em 350}, 17, doi:10.1016/j.physleta.2005.09.068.
\bibitem{Schutzhold2006} Sch\"{u}tzhold, R.; Schaller, G.; Habs, D. Signatures of the Unruh Effect from Electrons Accelerated by Ultrastrong Laser Fields. {\em Phys. Rev. Lett.} {\bf 2006}, {\em 97}, 121302, doi:10.1103/PhysRevLett.97.121302.
\bibitem{Retzker2008} Retzker, A.; Cirac, J.I.; Plenio, M.B.; Reznik, B. Methods for Detecting Acceleration
Radiation in a Bose-Einstein Condensate. {\em Phys. Rev. Lett.} {\bf 2008}, {\em 101}, 110402, doi:10.1103/PhysRevLett.101.110402.
\bibitem{Vanzella2001} Vanzella, D.A.T.; Matsas, G.E.A. Decay of Accelerated Protons and the Existence of the Fulling-Davies-Unruh Effect. {\em Phys. Rev. Lett.} {\bf 2001}, {\em 87}, 151301, doi:10.1103/PhysRevLett.87.151301.
\bibitem{Martine2011} Martin-Martinez, E.; Fuentes, I.; Mann, R.B. Using Berry's phase to detect the Unruh effect at lower accelerations. {\em Phys. Rev. Lett.} {\bf 2011}, {\em 107}, 131301, doi:10.1103/PhysRevLett.107.131301.
\bibitem{Pena2014} Pe\~{n}a, I.; Sudarsky, D. On the Possibility of Measuring the Unruh Effect. {\em Found. Phys.} {\bf 2014}, {\em 44}, 689, doi:10.1007/s10701-014-9806-0.
\bibitem{Cozzella2017} Cozzella, G.; Landulfo, A.G.S.; Matsas, G.E.A.; Vanzella, D.A.T. Proposal for Observing the Unruh Effect using Classical Electrodynamics. {\em Phys. Rev. Lett.} {\bf 2017}, {\em 118}, 161102, doi:10.1103/PhysRevLett.118.161102.
\bibitem{Matsas2002} Matsas, G.E.A.; Vanzella, D.A.T. The Fulling-Davies-Unruh effect is mandatory: The proton`s testimony. {\em Int.~J. Mod. Phys. D} {\bf 2002}, {\em 11}, 1573, doi:10.1142/S0218271802002918.
\bibitem{Audretsch1994} Audretsch, J.; M\"{u}ller, R. Spontaneous excitation of an accelerated atom: The contributions of vacuum fluctuations and radiation reaction. {\em Phys. Rev. A} {\bf 1994}, {\em 50}, 1755, doi:10.1103/PhysRevA.50.1755.
\bibitem{Audretsch1995} Audretsch, J.; M\"{u}ller, R. Radiative energy shifts of an accelerated two-level system. {\em Phys. Rev. A} {\bf 1995}, {\em 52}, 629, doi:10.1103/PhysRevA.52.629.
\bibitem{Passante1998} Passante, R. Radiative level shifts of an accelerated hydrogen atom and the Unruh effect in quantum electrodynamics. {\em Phys. Rev. A} {\bf 1998}, {\em 57}, 1590, doi:10.1103/PhysRevA.57.1590.
\bibitem{Zhu2006} Zhu, Z.; Yu, H.; Lu, S. Spontaneous excitation of an accelerated hydrogen atom coupled with electromagnetic vacuum fluctuations. {\em Phys. Rev. D} {\bf 2006}, {\em 73}, 107501, doi:10.1103/PhysRevD.73.107501.
\bibitem{Rizzuto2011} Rizzuto, L.;  Spagnolo, S. Energy-level shifts of a uniformly accelerated atom between two reflecting plates. {\em Phys. Scr.} {\bf 2011}, {\em T143}, 014021, doi:1088/0031-8949/2011/T143/014021.
\bibitem{Rizzuto2016} Rizzuto, L.; Lattuca, M.; Marino, J.; Noto, A.; Spagnolo, S.; Zhou, W.; Passante, R. Nonthermal effects of acceleration in the resonance interaction between two uniformly accelerated atoms. {\em Phys. Rev. A} {\bf 2016}, {\em 94}, 012121, doi:10.1103/PhysRevA.94.012121.
\bibitem{Lattuca2017} Lattuca, M.; Marino, J.; Noto, A.; Passante, R.; Rizzuto, L.; Spagnolo, S.; Zhou, W. Van der Waals and resonance interactions between accelerated atoms in vacuum and the Unruh effect. {\em J. Phys. Conf. Ser.} {\bf 2017}, {\em 880}, 012042, doi:10.1088/1742-6596/880/1/012042.
\bibitem{Yu2006} Yu, H.; Zhu, Z. Spontaneous absorption of an accelerated hydrogen atom near a conducting plane in vacuum. {\em Phys. Rev. D} {\bf 2006}, {\em 74}, 044032, doi:10.1103/PhysRevD.74.044032.
\bibitem{Zhu2007} Zhu, Z.; Yu, H. Fulling-Davies-Unruh effect and spontaneous excitation of an accelerated atom interacting with a quantum scalar field. {\em Phys. Lett. B} {\bf 2007}, {\em 645}, 459, doi:10.1016/j.physletb.2006.12.068.
\bibitem{Zhou2012} Zhou, W.; Yu, H. Spontaneous excitation of a uniformly accelerated atom coupled to vacuum Dirac field fluctuations. {\em Phys. Rev. A} {\bf 2012}, {\em 86}, 033841, doi:101103/PhysRevA.86.033841.
\bibitem{Rizzuto2007} Rizzuto, L. Casimir-Polder interaction between an accelerated two-level system and an infinite plate. {\em Phys. Rev. A} {\bf 2007}, {\em 76}, 062114, doi:10.1103/PhysRevA.76.062114.
\bibitem{Rizzuto2009a} Rizzuto, L.; Spagnolo, S. Lamb shift of a uniformly accelerated hydrogen atom in the presence of a conducting plate. {\em Phys. Rev. A} {\bf 2009}, {\em 79}, 062110, doi:10.1103/PhysRevA.79.062110.
\bibitem{Rizzuto2009b} Rizzuto, L.; Spagnolo, S. Energy level shifts of a uniformly accelerated atom in the presence of boundary conditions. {\em J. Phys. Conf. Ser.} {\bf 2009}, {\em 161}, 012031, doi:10.1088/1742-6596/161/1/012031.
\bibitem{Zhu2010} Zhu, Z.; Yu, H. Position-dependent energy-level shifts of an accelerated atom in the presence of a boundary. {\em Phys. Rev. A} {\bf 2010}, {\em 82}, 042108, doi:10.1103/PhysRevA.82.042108.
\bibitem{She2010} She, W.; Yu, H.; Zhu, Z. Casimir-Polder interaction between an atom and an infinite boundary in a thermal bath. {\em Phys. Rev. A} {\bf 2010}, {\em 81}, 012108, doi:10.1103/PhysRevA.81.012108.
\bibitem{Noto2013} Noto, A.; Passante, R. Van der Waals interaction energy between two atoms moving with uniform acceleration. {\em Phys. Rev. D} {\bf 2013}, {\em 88}, 025041, doi:10.1103/PhysRevD.88.025041.
\bibitem{Marino2014} Marino, J.; Noto, A.; Passante, R. Thermal and Nonthermal Signatures of the Unruh Effect in Casimir-Polder Forces. {\em Phys. Rev. Lett.} {\bf 2014}, {\em 113}, 020403, doi:10.1103/PhysRevLett.113.020403.
\bibitem{Antezza14} Antezza, M.; Braggio, C.; Carugno, G.; Noto, A.; Passante, R.; Rizzuto, L.; Ruoso, G.; Spagnolo, S. Optomechanical Rydberg-atom excitation via dynamic Casimir-Polder coupling. {\em Phys. Rev. Lett.} {\bf 2014}, {\em 113}, 023601, doi:10.1103/PhysRevLett.113.023601.
\bibitem{Bagarello15} Bagarello, F.; Lattuca, M.; Passante, R.; Rizzuto, L.; Spagnolo, S. Non-Hermitian Hamiltonian for a modulated Jaynes-Cummings model with ${\mathcal P T}$ symmetry. {\em Phys. Rev. A} {\bf 2015}, {\em 91}, 042134, doi:10.1103/PhysRevA.91.042134.
\bibitem{Zhou2016} Zhou, W.; Passante, R.; Rizzuto, L. Resonance interaction energy between two accelerated identical atoms in a coaccelerated frame and the Unruh effect. {\em Phys. Rev. D} {\bf 2016}, {\em 94}, 105025, doi:10.1103/PhysRevD.94.105025.
\bibitem{Power1982} Power, E.A.; Thirunamachandran, T. Quantum electrodynamics in a cavity. {\em Phys. Rev. A} {\bf 1982}, {\em 25}, 2473, doi:10.1103/PhysRevA.25.2473.
\bibitem{Meschede1990} Meschede, D.; Jhe, W.; Hinds, E.A. Radiative properties of atoms near a conducting plane: An old problem in a new light. {\em Phys. Rev. A} {\bf 1990}, {\em 41}, 1587, doi:10.1103/PhysRevA.41.1587.
\bibitem{Spagnolo2006} Spagnolo, S.; Passante, R.; Rizzuto, L. Field fluctuations near a conducting plate and Casimir-Polder forces in the presence of boundary conditions. {\em Phys. Rev. A} {\bf 2006}, {\em 73}, 062117, doi:10.1103/PhysRevA.73.062117.
\bibitem{Passante2007} Passante, R.; Spagnolo, S. Casimir-Polder interatomic potential between two atoms at finite temperature and in the presence of boundary conditions. {\em Phys. Rev. A} {\bf 2007}, {\em 76}, 042112, doi:10.1103/PhysRevA.76.042112.
\bibitem{Zhou2018} Zhou, W.; Rizzuto, L.; Passante, R. Vacuum fluctuations and radiation reaction contributions to the resonance dipole-dipole interaction between two atoms near a reflecting boundary. {\em Phys. Rev. A} {\bf 2018}, {\em 97}, 042503, doi:10.1103/PhysRevA.97.042503.
\bibitem{Palacino2017} Palacino, R.; Passante, R.; Rizzuto, R.; Barcellona, P.; Buhmann, S.Y. Tuning the collective decay of two entangled emitters by means of a nearby surface. {\em J. Phys. B At. Mol. Opt. Phys.} {\bf 2017}, {\em 50}, 154001, doi:10.1088/1361-6455/aa75f4.
\bibitem{Casimir1948} Casimir, H.B.G.; Polder, D. The Influence of Retardation on the London-van der Waals Forces. {\em Phys. Rev.} {\bf 1948}, {\em 73}, 360, doi:10.1103/PhysRev.73.360.
\bibitem{Salam10} Salam, A.  {\em Molecular Quantum Electrodynamics: Long-Range Intermolecular Interactions}; Wiley: Hoboken, NJ, USA, 2010.
\bibitem{Compagno1995} Compagno, G.; Passante, R.; Persico, F.  {\em Atom-Field Interactions and Dressed Atoms}; Cambridge University Press: Cambridge, UK, 1995.
\bibitem{Rizzuto2007a} Rizzuto, L.; Passante, R.; Persico, F. Nonlocal Properties of Dynamical Three-Body Casimir-Polder Forces. {\em Phys. Rev. Lett.} {\bf 2007}, {\em 98}, 240404, doi:10.1103/PhysRevLett.98.240404.
\bibitem{Rizzuto2004} Rizzuto, L.; Passante, R.; Persico, F. Dynamical Casimir-Polder energy between an excited- and a ground-state atom. {\em Phys. Rev. A} {\bf 2004}, {\em 70}, 012107, doi:10.1103/PhysRevA.70.012107.
\bibitem{Berman2015} Berman, P.R. Interaction energy of nonidentical atoms. {\em Phys. Rev. A} {\bf 2015}, {\em 91}, 042127, doi:10.1103/PhysRevA.91.042127.
\bibitem{Donaire2015} Donaire, M.; Guerout, R.; Lambrecht, A. Quasiresonant van der Waals Interaction between Nonidentical Atoms. {\em Phys. Rev. Lett.} {\bf 2015}, {\em 115}, 033201, doi:10.1103/PhysRevLett.115.033201.
\bibitem{Barcellona2016} Barcellona, P.; Passante, R.; Rizzuto, L.; Buhmann, S.Y. Van der Waals interactions between excited atoms in generic environments. {\em Phys. Rev. A} {\bf 2016}, {\em 94}, 012705, doi:10.1103/PhysRevA.94.012705.
\bibitem{Milonni2015} Milonni, P.W.; Rafsanjani, S.M.H. Distance dependence of two-atom dipole interactions with one atom in an excited state. {\em Phys. Rev. A} {\bf 2015} {\em 92}, 062711 doi:10.1103/PhysRevA.92.062711.
\bibitem{Forster1965} F\"{o}rster, T. {\em Modern Quantum Chemistry}; Academic: New York,  NY, USA, 1965.
\bibitem{Juzelinuas2000} Juzeli\={u}nas, G.; Andrews, D.L. Quantum Electrodynamics of Resonance Energy Transfer. {\em Adv. Chem. Phys.} {\bf 2000}, {\em 112}, 357, doi:10.1002/9780470141717.ch4.
\bibitem{Kurizki1996} Kurizki, G.; Kofman, A.G.; Yudson, V. Resonant photon exchange by atom pairs in high-Q cavities. {\em Phys. Rev. A} {\bf 1996}, {\em 53}, R35, doi:10.1103/PhysRevA.53.R35.
\bibitem{Agarwal1998} Agarwal, G.S.; Gupta, S.D. Microcavity-induced modification of the dipole-dipole interaction. {\em Phys. Rev. A} {\bf 1998}, {\em 57}, 667, doi:10.1103/PhysRevA.57.667.
\bibitem{Shahmoon2013} Shahmoon, E.; Kurizki, G. Nonradiative interaction and entanglement between distant atoms. {\em Phys. Rev. A} {\bf 2013}, {\em 87}, 033831, doi:10.1103/PhysRevA.87.033831.
\bibitem{Incardone2014} Incardone, R.; Fukuta, T.; Tanaka, S.; Petrosky, T.; Rizzuto, L.; Passante, R. Enhanced resonant force between two entangled identical atoms in a photonic crystal. {\em Phys. Rev. A} {\bf 2014}, {\em 89}, 062117, doi:10.1103/PhysRevA.89.062117.
\bibitem{Notararigo2018} Notararigo, V.; Passante, R.; Rizzuto, L. Resonance interaction energy between two entangled atoms in a photonic bandgap environment. {\em Sci. Rep.} {\bf 2018}, {\em 8}, 5193, doi:10.1038/s41598-018-23416-0.
\bibitem{Dalibard1982} Dalibard, J.; Dupont-Roc, J.; Cohen-Tannoudji, C. Vacuum fluctuations and radiation reaction: Identification of their respective contributions. {\em J. Phys.} {\bf 1982}, {\em 43}, 1617, doi:10.1051/jphys:0198200430110161700.
\bibitem{Dalibard1984} Dalibard, J.; Dupont-Roc, J.; Cohen-Tannoudji, C. Dynamics of a small system coupled to a reservoir: Reservoir fluctuations and self-reaction. {\em J. Phys.} {\bf 1984}, {\em 45}, 637, doi:10.1051/jphys:01984004504063700.
\bibitem{Menezes2016} Menezes, G.; Svaiter, N.F. Radiative processes of uniformly accelerated entangled atoms. {\em Phys. Rev. A} {\bf 2016}, {\em 93}, 052117, doi:10.1103/PhysRevA.93.052117.
\bibitem{Menezes2015} Menezes, G.; Svaiter, N.F. Vacuum fluctuations and radiation reaction in radiative processes of entangled states. {\em Phys. Rev. A} {\bf 2015}, {\em 92}, 062131, doi:10.1103/PhysRevA.92.062131.
\bibitem{ZhouYu2018} Zhou, W.; Yu, H. Boundarylike behaviors of the resonance interatomic energy in a cosmic string spacetime. {\em Phys. Rev. D} {\bf 2018}, {\em 97}, 045007, doi:10.1103/PhysRevD.97.045007.
\bibitem{Craig1998} Craig, D.P.; Thirunamachandran, T.  {\em Molecular Quantum Electrodynamics}; Dover Publ., Mineola, NY, USA,~1998.
\bibitem{Takagi1988} Takagi, S. Vacuum Noise and Stress Induced by Uniform Acceleration: Hawking-Unruh Effect in Rindler Manifold of Arbitrary Dimension. {\em Prog. Theor. Phys. Suppl.} {\bf 1988}, {\em 88}, 1, doi:10.1143/PTP.88.1.
\end{thebibliography}
\end{document}